\documentclass[trackchanges,twocolumn]{aastex701}

\begin{document}

   \title{Transformers for Stratified Spectropolarimetric Inversion: Proof of Concept} 

\author[orcid=0000-0001-5699-2991,sname='Campbell', gname=Ryan]{Ryan J. Campbell}

\affiliation{Astrophysics Research Centre, Queen's University Belfast, Belfast, Northern Ireland, BT7 1NN, United Kingdom}
\email[show]{rjcampbell.research@outlook.com}  
\correspondingauthor{R. J. Campbell}

\author[orcid=0000-0002-7725-6296,sname='Mathioudakis', gname=Mihalis]{Mihalis Mathioudakis}
\affiliation{Astrophysics Research Centre, Queen's University Belfast, Belfast, Northern Ireland, BT7 1NN, United Kingdom}
\email{m.mathioudakis@qub.ac.uk} 

\author[orcid=0000-0001-5518-8782, gname=Carlos,sname=Quintero Noda]{Carlos Quintero Noda}
\affiliation{Instituto de Astrof\'isica de Canarias,
V\'ia L\'actea s/n, E-38205 La Laguna,
Tenerife, Spain}
\affiliation{Departamento de Astrof\'isica, Universidad de La Laguna, E-38206 La Laguna, Tenerife, Spain}
\email{carlos.quintero@iac.es}

\begin{abstract}
Solar spectropolarimetric inversion---inferring atmospheric conditions from the Stokes vector---is a key diagnostic tool for understanding solar magnetism, but traditional inversion methods are computationally expensive and sensitive to local minima. Advances in artificial intelligence (AI) offer faster solutions, but are often restricted to shallow models or a few spectral lines. We present a proof-of-concept study using a transformer machine learning (ML) model for multi-line, full-Stokes inversion, to infer stratified parameters from synthetic spectra produced from 3D magnetohydrodynamic simulations. We synthesise a large set of Stokes vectors using forward modelling across 15 spectral lines spanning the deep photosphere towards the chromosphere. The model maps full-Stokes input to temperature, magnetic field strength, inclination, azimuth (encoded as $\sin2\phi$, $\cos2\phi$), and line-of-sight velocity as a function of optical depth. The transformer incorporates an attention mechanism that allows the model to focus on the most informative regions of the spectrum for each inferred parameter, and uses positional embedding to encode wavelength and depth order. We benchmark it against a multilayer perceptron (MLP), test robustness to noise, and assess generalisation. The transformer outperforms the MLP, especially in the higher layers and for magnetic parameters, yielding higher correlations and more regularised stratifications. The model retains strong performance across a range of noise levels typical for real observations, with magnetic parameter inference degrading predictably while temperature and velocity remain stable. We explore attention maps, linking the transformer's learned behaviour to line-formation physics.
\end{abstract}


\section{Introduction}
Inverting the Stokes vector to retrieve physical conditions in the solar atmosphere is fundamental for understanding solar magnetism \citep{inversions2016}. Inversion is the inference of depth-dependent atmospheric parameters from the wavelength dependence of the polarized spectrum in spectral lines and corresponding continua. While traditional inversion methods based on radiative transfer are physically grounded, they are also computationally intensive and sensitive to initialisation and noise \citep{ramos2012}. Crucially, the inversion problem is well suited to machine learning (ML): it can be framed as a supervised regression task with known forward physics, allowing large synthetic training sets with perfect labels, and a deterministic mapping from input (Stokes vectors as a function of wavelength) to output (atmospheric parameters as a function of geometric height or optical depth). These properties make spectropolarimetric inversion an ideal application for ML architectures, especially those that can handle high-dimensional inputs and uncertain mappings.

ML has been widely applied in solar physics, particularly to speed up the retrieval of atmospheric parameters from observed Stokes vectors \citep{Ramos2019, RamosLRSP}. A common approach involves using inversions of a small subset of Stokes vectors and associated atmospheric models as training data for a Multi-Layer Perceptron (MLP) or Convolutional Neural Network (CNN), which can then rapidly provide atmospheric models for the rest of the dataset \citep{Milic2020}. However, a major challenge is providing a diverse enough training set to achieve sufficient generalisation, necessitating large data volumes \citep{SPIn4D}. \citet{Gafeira2021} improved the accuracy of their inversions by using models produced from CNNs as initial conditions for traditional inversions. \citet{Navarro} used only forward synthesis, an MLP, and randomly generated atmospheric models to infer target parameters, demonstrating generalisation to unseen Hinode data. Another recent advance is the use of neural networks influenced by physical equations, such as the approach taken by \citet{PINNME}, \cite{keller}, or \cite{Ramos2025}.

Transformers, originally developed for natural language processing \citep{vaswani2017attention}, have demonstrated success across many domains due to their ability to model long-range dependencies and flexible input structures. By applying a transformer-based architecture to the spectropolarimetric inversion problem, we aim to capture subtle spectral features and inter-wavelength correlations.

\subsection{Aims and hypothesis} Our workflow comprises the following steps:
\begin{enumerate}
      \item Synthesise a large  training set of Stokes vectors from 3D MHD simulations using a suitable radiative transfer code.
      \item Implement a transformer model that takes full Stokes vectors as input and outputs stratified atmospheric parameters (e.g., temperature, magnetic field vector, line-of-sight velocity) as a function of optical depth.
      \item Assess the model’s performance using synthetic data, evaluating its ability to generalise, capture vertical structure, and benchmark against an existing ML sequence-based method.
\end{enumerate}

The move from MLPs or CNNs to transformers fundamentally changes what the model can learn. Transformers allow the model to attend to non-local spectral features across the ordered line profile, which is especially important in spectropolarimetry where physical information (like Zeeman splitting or velocity gradients) can manifest in subtle, non-local ways. By ``non-local'', we refer to correlations or informative relationships between widely separated wavelength points. Unlike MLPs, which treat each wavelength as an independent input feature and do not have awareness of their order, transformers can dynamically learn the spectral context. By ``context'', we mean features whose interpretation depends on the broader spectral profiles. To our knowledge, at the time of submission, this was the first application of an attention-based (transformer) neural network to the problem of spectropolarimetric inversion in solar physics. Shortly following submission, \cite{ramos_jaime_2025} applied a transformer architecture to the inversion problem with a neural model used to compress the data into a lower dimensionality. Our central hypothesis is that a transformer-based architecture will achieve more accurate spectropolarimetric inversions than an MLP baseline, due to its ability to model spectral order (via learned positional embeddings) and non-local spectral dependencies (via attention mechanisms). For our study, we choose to benchmark the transformer to an MLP because without an attention mechanism the transformer architecture reduces, albeit with some caveats, to a very deep stack of MLPs (i.e. feed-forward networks). This allows us to isolate the impact of the attention mechanism and positional embeddings on inversion performance most effectively and makes a well-tuned MLP the most natural baseline model.


\section{Data}
\subsection{Simulation snapshots}
For this study, we employ three-dimensional snapshots generated with the MANCHA code \citep{MANCHA1}, as described by \cite{MANCHA2}. By choosing a quiet Sun simulation, we deliberately select one of the most challenging regimes for spectropolarimetric inversion, where polarisation signals, particularly in the chromosphere, are weak. The simulation starts from an initially field-free state, allowing the Biermann battery effect and a local turbulent dynamo to self-consistently generate magnetic fields reaching strengths comparable to those observed in the quiet Sun. These simulation snapshots have also been used in previous works such as \cite{Carlos2023,Campbell2025}. The computational domain covers an area of $11.52 \times 11.52$~Mm$^2$ (resolved with $1152 \times 1152$~pixels$^2$), and extends vertically from $1$~Mm beneath the photosphere, with a $1400$~km upper limit. The grid spacing is $10$~km in the horizontal directions and $7$~km vertically. The mean surface magnetic field strength in the saturated regime is approximately $100$~G.

To assess the generalisation performance of the ML models, we train on a given snapshot and evaluate inference on a different snapshot obtained $450$ seconds later, thereby guaranteeing that the evaluation is performed on physically independent data not encountered during training. 

While the chromospheric layers in the MANCHA simulation may not fully reflect the complexity of real solar conditions, they still influence the formation of spectral lines such as Ca II $854$ nm, allowing the model to learn meaningful mappings between Stokes profiles and atmospheric parameters in those layers.

\subsection{Forward synthesis}
To generate the synthetic observables for this study, we use the DeSIRe code \citep{DeSIRe} to forward-synthesise a set of spectral lines covering atmospheric heights from the deep photosphere to the chromosphere, enabling a comprehensive assessment of model performance at different optical depths.

The simulation outputs are provided on a three-dimensional grid in geometrical height (km). To facilitate forward synthesis with the DeSIRe code, which requires an optical depth scale, we converted the atmospheric stratifications to the required $\log(\tau)$ grid. This was done using the \textit{modelador.x} utility distributed with DeSIRe, following the methodology outlined in \cite{Campbell2021}. The final grid has $81$ depth points.

\begin{figure*}
    \includegraphics[width=\textwidth]{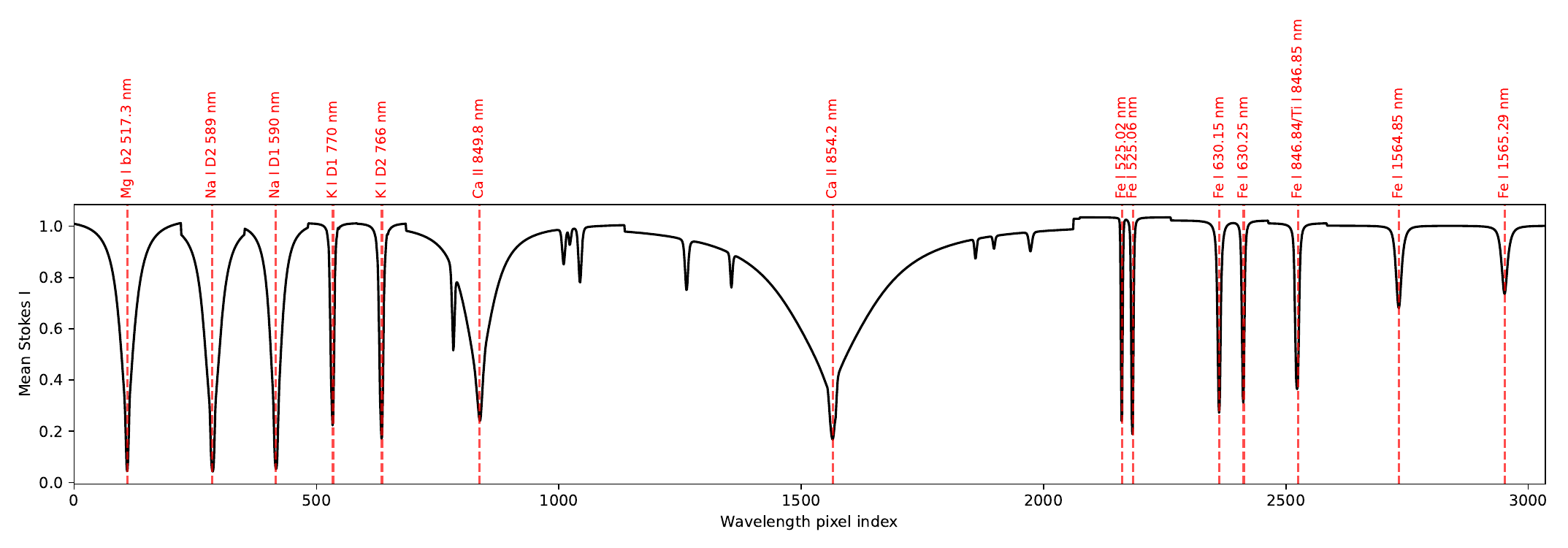}
    \caption{Mean Stokes $I$ profile across the full field of view, averaged over all spatial pixels in the MANCHA simulation snapshot used for training. The vertical dashed lines and labels indicate the most important diagnostic spectral lines. These lines span a broad range of atmospheric heights, from the deep photosphere to the mid-chromosphere, providing constraints for inversion at multiple depths.}
    \label{fig:mean_intensity}
\end{figure*}

Figure~\ref{fig:mean_intensity} shows the mean intensity profile across a full snapshot. The diagnostic lines we synthesised are shown in Table~\ref{table:lines}. The Fe I $525$~nm, $630$~nm, and $1565$~nm line pairs, and the Fe I $846.84$ nm line, are synthesised in local thermodynamic equilibrium (LTE); all others are computed in non-LTE (NLTE). All spectral lines are synthesised with a spectral sampling of $20$~m$\mathrm{\AA}$ per pixel. All lines were synthesised under the assumption of complete redistribution (CRD) when relevant. The spectral lines were synthesised with an observation angle, $\mu$, of $1.0$.

\begin{table*}
\begin{tabular}{lllll}
\hline
Spectral line & $\log gf$ & $E_{\rm low}$ (eV) & Transition & Sensitivity \\
\hline
Mg I b2 517.3 nm$^\ast$        & $-0.38616$   & 2.7166  & $3P_{1.0} - 3S_{1.0}$    & Upper photosphere/lower chromosphere \\
Na I D2 589 nm$^\ast$           & $+0.10106$   & 0.000   & $2S_{0.5} - 2P_{1.5}$    & Upper photosphere    \\
Na I D1 590 nm$^\ast$           & $-0.18402$   & 0.000   & $2S_{0.5} - 2P_{0.5}$    & Upper photosphere    \\
K I D1 770 nm$^\ast$            & $-0.16876$   & 0.000   & $2S_{0.5} - 2P_{0.5}$    & Upper photosphere    \\
K I D2 766 nm$^\ast$            & $+0.1345$    & 0.000   & $2S_{0.5} - 2P_{1.5}$    & Upper photosphere    \\
Ca II 849.8 nm$^\ast$           & $-1.31194$   & 1.6924  & $2D_{1.5} - 2P_{1.5}$    & Mid chromosphere (core)    \\
Ca II 854.2 nm$^\ast$           & $-0.36199$   & 1.7000  & $2D_{2.5} - 2P_{1.5}$    & Mid chromosphere (core)    \\
Fe I 525.02 nm           & $-4.938$     & 0.121   & $5D_{0.0} - 7D_{1.0}$    & Photosphere          \\
Fe I 525.06 nm           & $-2.181$     & 2.198   & $5P_{2.0} - 5P_{3.0}$    & Photosphere          \\
Fe I 630.15 nm           & $-0.718$     & 3.654   & $5P_{2.0} - 5D_{2.0}$    & Photosphere          \\
Fe I 630.25 nm           & $-1.131$     & 3.686   & $5P_{1.0} - 5D_{0.0}$    & Photosphere          \\
Fe I 846.84 nm           & $-2.072$     & 2.221   & $5P_{1.0} - 5P_{1.0}$    & Photosphere          \\
Ti I 846.85 nm           & $-0.993$     & 1.888   & $3G_{5.0} - 3F_{4.0}$    & Photosphere          \\
Fe I 1564.85 nm          & $-0.652$     & 5.426   & $7D_{1.0} - 7D_{1.0}$    & Deep photosphere     \\
Fe I 1565.29 nm          & $+0.001$     & 6.246   & $7D_{5.0} - (6D_{4.5})f2k_{4.0}$ & Photosphere  \\
\hline
\end{tabular}
\caption{Atomic data and dominant formation layer for the selected spectral lines. “Sensitivity” is approximate and assumes quiet Sun conditions. Blended lines in the Ca II lines are not listed. The $^\ast$ indicates that the spectral line was synthesised in NLTE.}
\label{table:lines}
\end{table*}

\section{Methods}
\subsection{Preprocessing}
Before training, all models undergo a common preprocessing pipeline applied to both inputs and outputs. These steps ensure consistent data formatting, scaling, and physical representation across architectures. All input and output variables (excluding azimuth) are standardised using the global mean, $\mu$, and standard deviation, $\sigma$, computed from the training set:
\begin{equation}
x_{\text{scaled}} = \frac{x - \mu}{\sigma}.
\end{equation}
The azimuth is instead represented using $\sin(2\phi)$ and $\cos(2\phi)$, which are bounded between –1 and 1 and therefore do not require standardisation. After inference, the model outputs are inverse-transformed to physical units using the same training-set statistics.

\subsection{Transformer architecture}
\begin{figure}[h]
\centering
\includegraphics[width=\columnwidth]{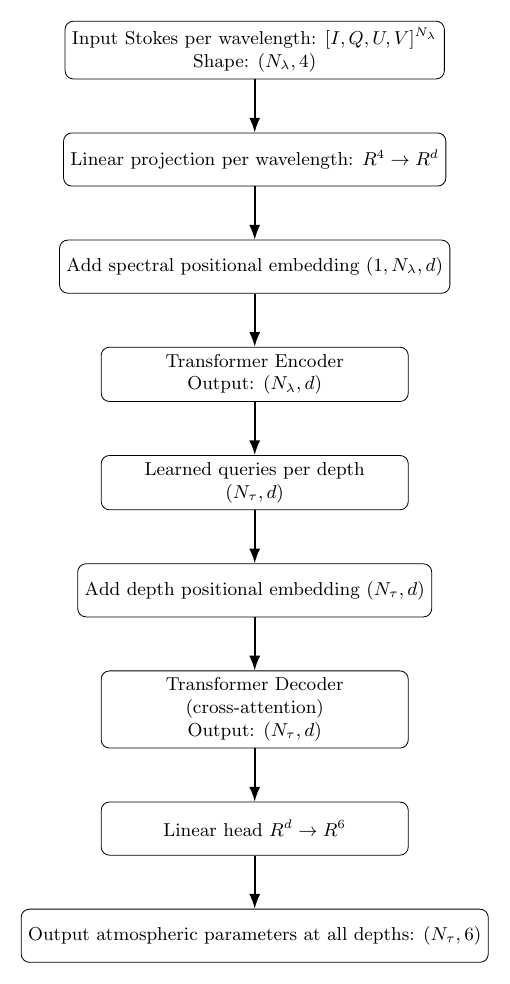}
\caption{Flow diagram of the transformer-based inversion model. The encoder processes sequences of Stokes vectors across wavelength positions, where each token corresponds to a single wavelength and contains the full 4-dimensional Stokes vector. The decoder uses learned queries for each depth point to extract atmospheric parameters via cross-attention.}
\label{fig:transformer_architecture}
\end{figure}

We implement a transformer-based neural network to perform spectropolarimetric inversions, predicting physical parameters of the solar atmosphere from observed Stokes profiles. The model ingests full Stokes vectors across multiple wavelengths and outputs stratified atmospheric parameters at all optical depth points. The transformer architecture was originally developed for natural language processing (NLP) tasks, where it models sequences of tokens and learns the contextual meaning of each token based on its surrounding token. In that setting, attention mechanisms allow the model to focus on the most relevant parts of the sentence when generating or interpreting text. Analogously, in our application, each ``token'' corresponds to the Stokes vector at a particular wavelength. The attention mechanism allows the model to learn which parts of the spectral line are most informative for inferring each physical parameter. 

The transformer architecture is described by Fig.~\ref{fig:transformer_architecture}. Each input sequence corresponds to a spectral profile at a given spatial pixel, represented as a sequence of tokens. Each token corresponds to a single wavelength point and contains a 4-dimensional Stokes vector: $[I, Q, U, V]$. This forms a sequence of tokens with shape $(N_\lambda, 4)$. The model linearly projects each token into a higher-dimensional embedding space and adds learned spectral positional embeddings to preserve wavelength order. Otherwise, the model would treat the spectral sequence as unordered.\footnote{The term positional encoding is often used. We adopt the term learned positional embedding here to reflect that the spectral positional information is implemented as a learned parameter matrix, rather than a fixed encoding.}. During model development, we ran ablation tests with positional embedding disabled. We found that spectral positional embedding was essential, as model performance collapsed when it was removed. In contrast, depth positional embedding was less critical: the model achieved comparable validation loss without it. However, we retained it in the final model to promote depth-wise regularisation.

\begin{figure}[h]
\centering
\includegraphics[width=\columnwidth]{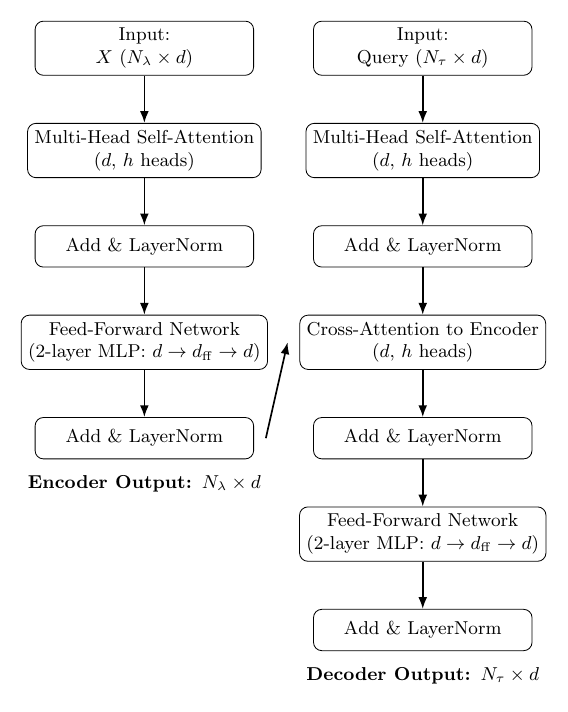}
\caption{Schematic diagram of a single transformer encoder and decoder layer. Each encoder layer consists of multi-head self-attention, a position-wise feed-forward network (MLP), and residual connections. Each decoder layer additionally includes cross-attention to the encoder output. $d$ is the model (embedding) dimension, $h$ is the number of heads, $d_\mathrm{ff}$ is the MLP hidden dimension. Add and LayerNorm adds the input of each transformer block (e.g. attention or feed-forward) to its output, and then applies normalisation to stabilise training; this is a standard architectural feature of modern transformer networks.}
\label{fig:enc-dec}
\end{figure}

We adopt an architectural design that enables a natural separation between the spectral domain (handled by an encoder) and the depth-dependent output domain (handled by a decoder), while preserving flexibility in the number and identity of inferred physical parameters. This encoder–decoder architecture is illustrated in Fig.~\ref{fig:enc-dec}.
The embedded sequence is processed by a transformer encoder composed of stacked multi-head self-attention layers and feed-forward networks. 
Our encoder and decoder each consist of a tunable number of stacked layers. Each layer contains a multi-head attention mechanism followed by a position-wise feed-forward network \footnote{“Position-wise” means that the same feed-forward network (a two-layer MLP) is applied independently to each token in the sequence. There is no interaction between sequence positions in this sublayer; all contextual mixing occurs in the attention mechanism. The feedforward sublayer (MLP) inside each transformer block uses a hidden size of $2048$ (PyTorch default).} (two linear layers with a dropout rate of 0.1 and ReLU activation\footnote{ReLU (Rectified Linear Unit) is a common activation function defined as $\mathrm{ReLU}(x) = \max(0, x)$. Activation functions introduce non-linearity into neural networks, enabling them to approximate complex, non-linear mappings between inputs and outputs. Without them, the entire model would collapse into a linear transformation, regardless of depth.}). These MLP sublayers are essential for learning non-linear combinations of spectral or depth-specific features at each token position. Each attention module uses a controllable number of heads, allowing the model to learn multiple distinct attention patterns in parallel. The outputs from each head are concatenated and linearly transformed to form the final representation for that layer.
The self-attention layers compute contextualised representations of each wavelength point by comparing it to all others, enabling the model to capture subtle relationships across the spectral line profile.

In the transformer framework, the core computation within each attention layer involves three components: queries, keys, and values, as originally introduced by \citet{vaswani2017attention}. For a given position in the sequence (e.g., a specific wavelength), the query vector determines what the model is looking for, the key vectors determine where that information is located in the input, and the value vectors represent what is retrieved. For a given input sequence $X \in \mathbb{R}^{N_\lambda \times d}$ after projection, the queries, keys, and values are computed as
\begin{equation}
Q = XW^Q, \quad K = XW^K, \quad V = XW^V,
\end{equation}
where $W^Q \in \mathbb{R}^{d \times d}$, and similarly for $W^K$ and $W^V$, are learned weight matrices, and the dimensionality, $d$, is a model hyperparameter that controls the capacity of the attention mechanism.

The self-attention mechanism then computes attention-weighted combinations of the values:
\begin{equation}\label{eqn:attn}
\text{Attention}(Q, K, V) = \text{softmax}\left( \frac{QK^\top}{\sqrt{d}} \right) V.
\end{equation}
The dot product $QK^\top$ quantifies the similarity between the query and each key, and the softmax function converts these similarities into a probability distribution used to weight the values. This enables each wavelength position to attend to the most relevant spectral features, learned through optimisation of the model parameters during training.
In other words, larger values of the dot product indicate greater relevance, allowing the model to attend more strongly to parts of the input sequence that are most aligned with the current query. The desired behaviour emerges because all weights—including those defining queries, keys, and values—are learned during training in a way that minimises a predefined loss function, ensuring that attention patterns are shaped to optimise the predictive accuracy of the model.

To generate predictions at multiple optical depth points, we adopt a decoder-based architecture. A fixed set of learnable query embeddings—one for each optical depth point—is passed to a transformer decoder. Each query is augmented with a learnable optical depth positional embedding, allowing the decoder to distinguish between physical layers. The decoder then uses multi-head cross-attention to attend to the encoder output, learning how each layer of the atmosphere relates to the observed spectral features. Decoder self-attention is retained, allowing depth layers to share information, which promotes vertical regularity and physical coherence in the inferred atmospheric profiles.

Thus, in summary, our model includes self-attention in the encoder (across wavelengths), cross-attention in the decoder (between depths and spectral features), and self-attention in the decoder (across depths). Each decoder layer performs cross-attention with the common encoder output irrespective of the number of encoder layers, ensuring that all depth queries attend to a shared spectral representation.

\subsection{MLP baseline}

\begin{figure}[h]
\centering
\includegraphics[width=\columnwidth]{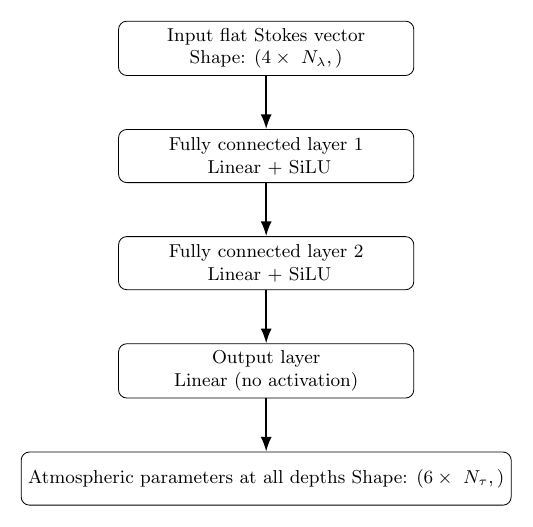}
\caption{Flow diagram of a 2-hidden-layer MLP inversion model. The entire input spectrum is first flattened and then passed through three fully connected layers, producing the full set of atmospheric parameters at all depths as output.}
\label{fig:MLP_architecture}
\end{figure}

To provide a simple and fast baseline, we also train a fully connected multilayer perceptron (MLP) to perform the same inversion task as the transformer. The input to the MLP is a flattened Stokes vector of shape \( (4 \times N_\lambda) \), and the output is a vector of size \( (6 \times N_\tau) \), representing all physical parameters at all optical depth points. As Figure~\ref{fig:MLP_architecture} illustrates, a 2-hidden-layer MLP architecture consists of three fully connected layers. Each hidden layer applies a linear transformation followed by a Sigmoid Linear Unit (SiLU) activation; the final output layer is linear. We use MLP architectures with more than two hidden layers, but Fig. $\ref{fig:MLP_architecture}$ shows an example with only two for clarity.

This architecture processes each pixel's Stokes vector in isolation, with no internal representation of spectral structure, positional embedding, or attention mechanism. The MLP treats the flattened input as an unordered list, lacking any explicit or implicit encoding of wavelength order, spectral continuity, or the ordering of optical depth points in the output. As a result, it cannot exploit non-local or context-dependent features in the data, and serves as a point of comparison for more expressive sequence-based models. While the MLP baseline trains significantly faster than the transformer model, it has no explicit mechanism to model relationships along the spectrum or across optical depths, relying instead on learning such dependencies implicitly from the data.

\subsection{Training and validation}
The final output for both models consists of six values per optical depth point: temperature, magnetic field strength, line-of-sight velocity, magnetic inclination, and the azimuth is represented using $\sin(2\phi)$ and $\cos(2\phi)$, which transforms the periodic azimuth angle into a continuous space that avoids discontinuities near $0^\circ, 180^\circ$, though this does not fully resolve the inherent ambiguity in the azimuthal direction.

For both the transformer and MLP models, we generate a total of 1 million unique training examples from a single snapshot of the MANCHA simulation, consisting of synthetic Stokes profiles and their corresponding atmospheric models. These data are randomly split into training and validation sets ($85\%/15\%$). The generalisation and test performance are evaluated on a completely independent snapshot from the same simulation, taken $450$~seconds later. Both the transformer and MLP are trained and evaluated on identical datasets to enable a direct comparison.

We learn all model parameters end-to-end via back-propagation of the mean squared error (MSE) loss, computed as the average squared difference over all outputs, depths, and batch elements between predicted and ground truth parameters:
\begin{equation}
\mathcal{L} = \frac{1}{N} \sum_{i=1}^{N} \left\| \hat{y}_i - y_i \right\|^2,
\end{equation}
where $\hat{y}_i$ and $y_i$ are the predicted and true outputs for the $i$-th training example, and $N$ is the number of examples in the batch. We experimented with both the Adam and AdamW optimizers \citep{Adam}. The latter includes a modest weight decay ($1\times10^{-4}$) to gently penalize overly large weights and improve generalization. We used mixed-precision (16-bit) arithmetic with automatic loss scaling, which speeds up training and reduces memory without sacrificing stability.  To prevent any single batch from causing a sudden, oversized update, we cap the combined size of all gradient values at $1.0$ so if the raw gradients exceed this threshold, they’re uniformly scaled down such that the model only makes a controlled adjustment. We performed early stopping based on validation loss, with a maximum of 35 epochs per run. 

Hyperparameter searches are performed for both architectures: for the transformer, we vary learning rate ({0.001, $1\times10^{-4}$, $3\times10^{-4}$}), hidden dimension ({128, 256, 384}), number of encoder and decoder layers ({2, 3, 4}), and number of attention heads ({2, 4}). For the MLP baseline, we explore learning rates ({0.001, $1\times10^{-4}$, $3\times10^{-4}$}), hidden dimensions ({128, 256, 384, 512, 1024, 2048}), and the number of layers ({3, 4, 6, 8}). The batch size is primarily determined by available GPU memory, with typical values of 32--64. The best model for each architecture is selected as that with the lowest validation loss during training. For the MLP, the lowest validation loss was achieved with a learning rate of $3\times10^{-4}$, a hidden dimension of 2048, and 8 layers. For the transformer, the lowest validation loss was achieved with a learning rate of $3\times10^{-4}$, a hidden dimension of 128, 4 layers, and 4 attention heads. 

We also considered the impact of dropout on the performance of the MLP. The transformer applies a dropout rate of $0.1$ within its feed-forward layers and residual connections, as per the default PyTorch implementation. We added dropout layers after each activation in the MLP architecture, using a dropout rate of $0.1$ to match the default setting in the Transformer's encoder and decoder blocks. The validation loss increased significantly, suggesting that dropout was not beneficial for the MLP in this setting.

Training was performed on NVIDIA H100 Tensor Core GPUs, a state-of-the-art hardware platform designed for large-scale deep learning workloads. The H100 is widely used for training large neural networks, including transformer architectures in applications such as natural language processing and computer vision. Training the transformer model on 1 million profiles (and associated models) required approximately 12 hours for 35 epochs on a single NVIDIA H100 GPU. For comparison, training the baseline MLP on the same dataset required approximately 30 minutes for 35 epochs. Inference on a full snapshot is performed much faster, enabling rapid processing of large spectropolarimetric datasets. The longer training time for the transformer reflects both its architectural complexity and the computational cost of the self-attention mechanism, which scales quadratically with input sequence length (i.e. the number of wavelengths). This is in contrast to the MLP, whose feed-forward layers scale linearly with input size. 

The code used in this project to train the machine learning models is open source and is available\footnote{\url{www.github.com/r-j-campbell/SINN-inversions}}\footnote{\url{https://doi.org/10.5281/zenodo.16793363}}.

\section{Results and analysis}
\subsection{Loss curves}
\begin{figure}
    \centering

    \includegraphics[width=1.0\linewidth]{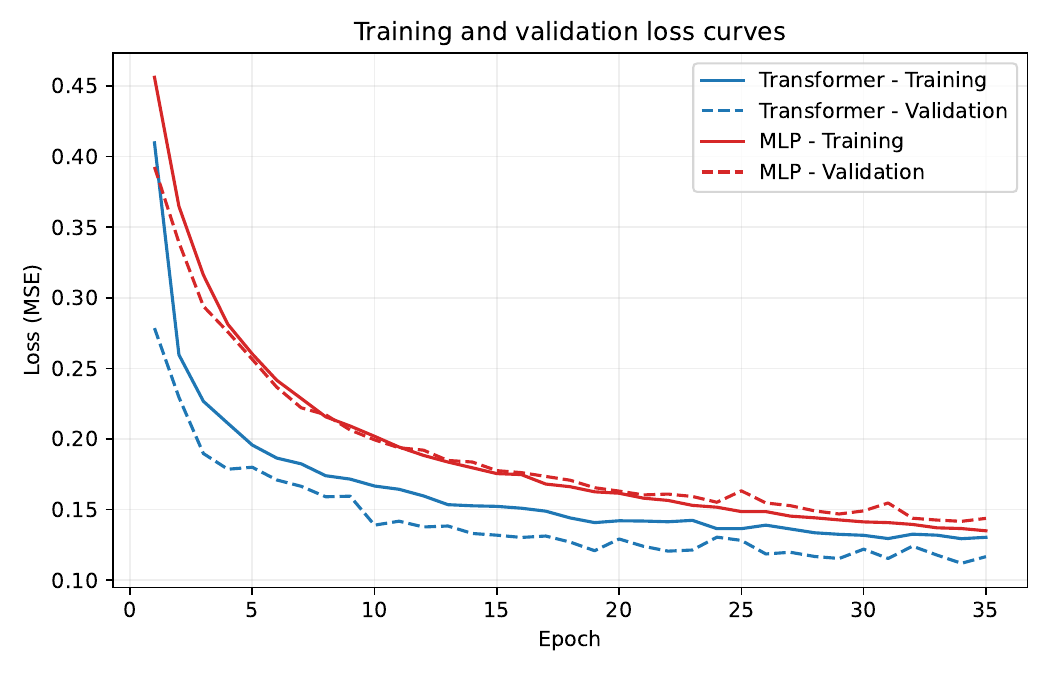}
    \caption{Training (solid lines) and validation (dashed lines) loss curves for the transformer (blue) and MLP (red) models, as a function of epoch. The epoch at which the transformer achieves its lowest validation loss is epoch 34 with a validation loss of 0.112. The lowest validation loss for the MLP was at epoch 34 with 0.1417.}
    \label{fig:loss_curves}
\end{figure}

Before benchmarking inversion performance, we first examine the training and validation loss curves for both the transformer and MLP models, shown in Figure~\ref{fig:loss_curves}. For each architecture, we plot the MSE on the training and validation sets as a function of epoch for the best-performing hyperparameter configuration. In both cases, the loss curves show smooth convergence, with validation loss plateauing, indicating effective early stopping and an absence of overfitting. This ensures that subsequent comparisons are made using well-generalised models.

In the MLP run, the close convergence of the training and validation loss curves, without a widening gap or upward trend in validation loss, indicates neither underfitting (where both losses would stagnate at high values) nor overfitting (where training loss would decrease while validation loss rises sharply). For the transformer, in the best run, the validation loss is observed to be consistently slightly lower than the training loss. This arises because, during training, loss is computed and averaged over individual mini-batches, which introduces some stochastic variability and can result in a slightly higher reported value. In contrast, the validation loss is calculated over the entire validation set at once, providing a more stable and, in some cases, lower value. This behaviour is common in large deep learning models and does not indicate underfitting or overfitting. The important diagnostic is that both losses converge and flatten without a widening gap.

\subsection{Generalisation testing and comparing the transformer and MLP predictions}
\begin{figure*}
    \centering
    \includegraphics[width=\textwidth]{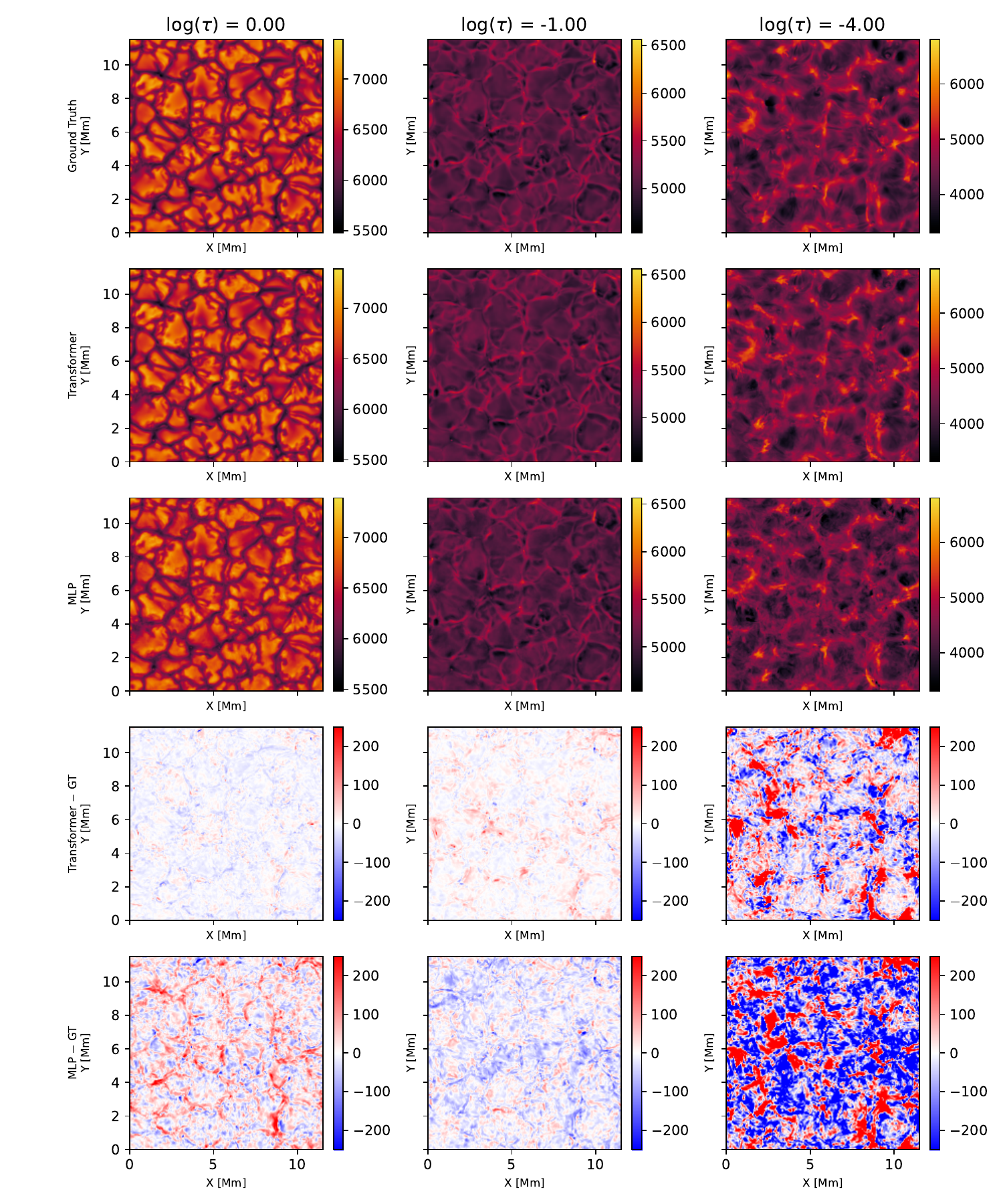}
    \caption{Temperature maps at selected optical depths comparing ground truth, transformer, and MLP predictions. The top row shows the ground truth temperature maps from the MANCHA simulation. The second and third rows display the corresponding predictions from the transformer and MLP models, respectively. The fourth and fifth rows show the residual maps (prediction minus ground truth) for the transformer and MLP models. Colorbars indicate temperatures in Kelvin for the absolute maps and differences in Kelvin for the residuals. }
    \label{fig:maps_T}
\end{figure*}

Figure~\ref{fig:maps_T} shows maps of the temperature at selected optical depths, comparing the ground truth, transformer prediction, MLP prediction, and the residuals. At $\log(\tau) = 0$, the transformer accurately reproduces the surface granulation pattern, capturing even subtle small-scale features within intergranular lanes. The MLP performs similarly well across most of the field of view, but exhibits larger errors than the transformer, particularly within intergranular lanes. The residual maps confirm that the transformer outperforms the MLP throughout, with the largest MLP errors exceeding 200 K, whereas transformer errors in the same regions remain typically below 200 K, reflecting greater stability. These trends persist at $\log(\tau) = -1$, with the transformer consistently delivering more accurate and robust predictions. However, at $\log(\tau) = -4$, both models show increased absolute residuals and systematic errors, though the transformer still better captures the overall temperature morphology.

\begin{figure*}
    \centering
    \includegraphics[width=\textwidth]{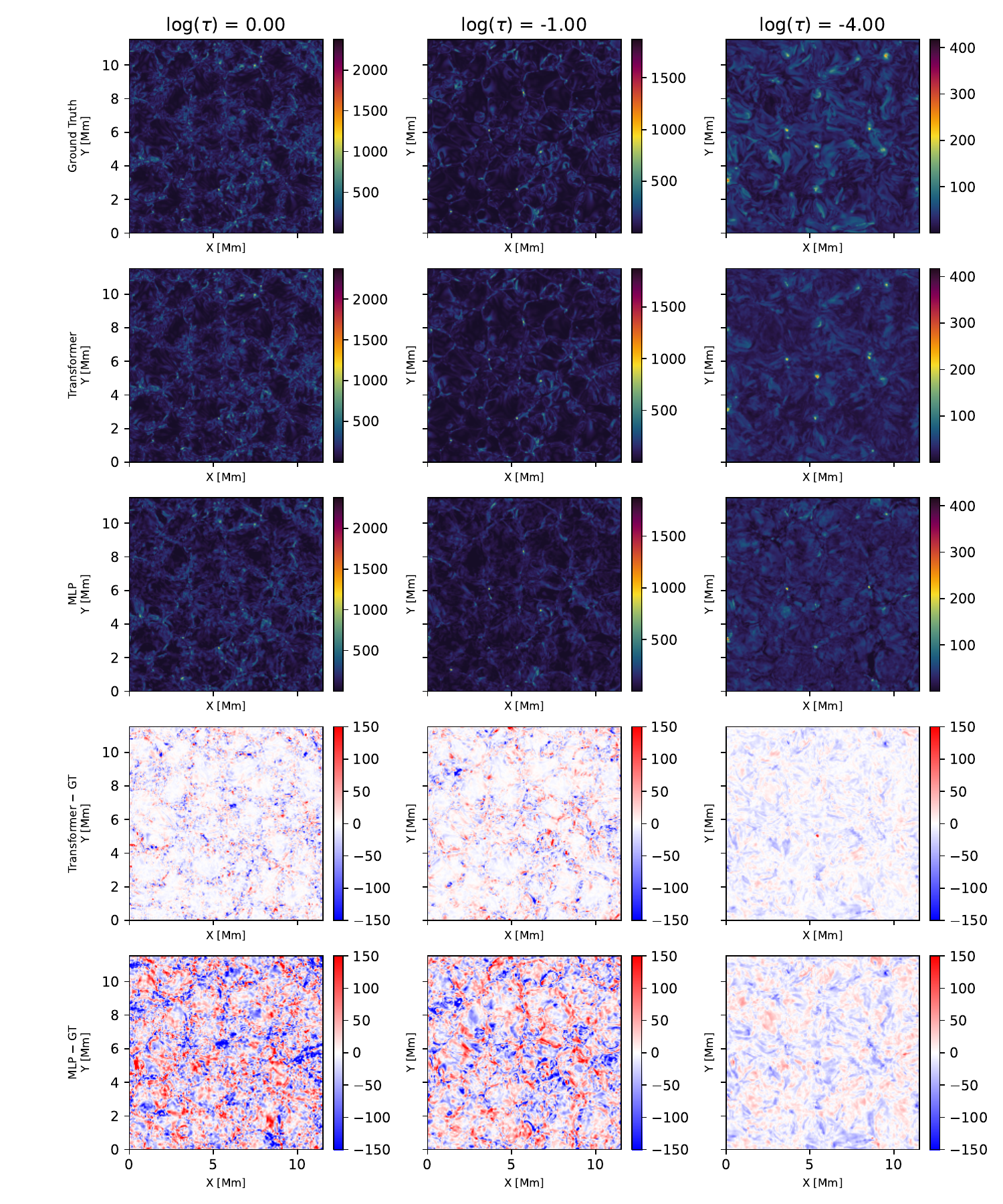}
    \caption{As in Fig.~\ref{fig:maps_T}, but for magnetic field strength. All units are Gauss.}
    \label{fig:maps_B}
\end{figure*}

\begin{figure*}
    \centering
    \includegraphics[width=\textwidth]{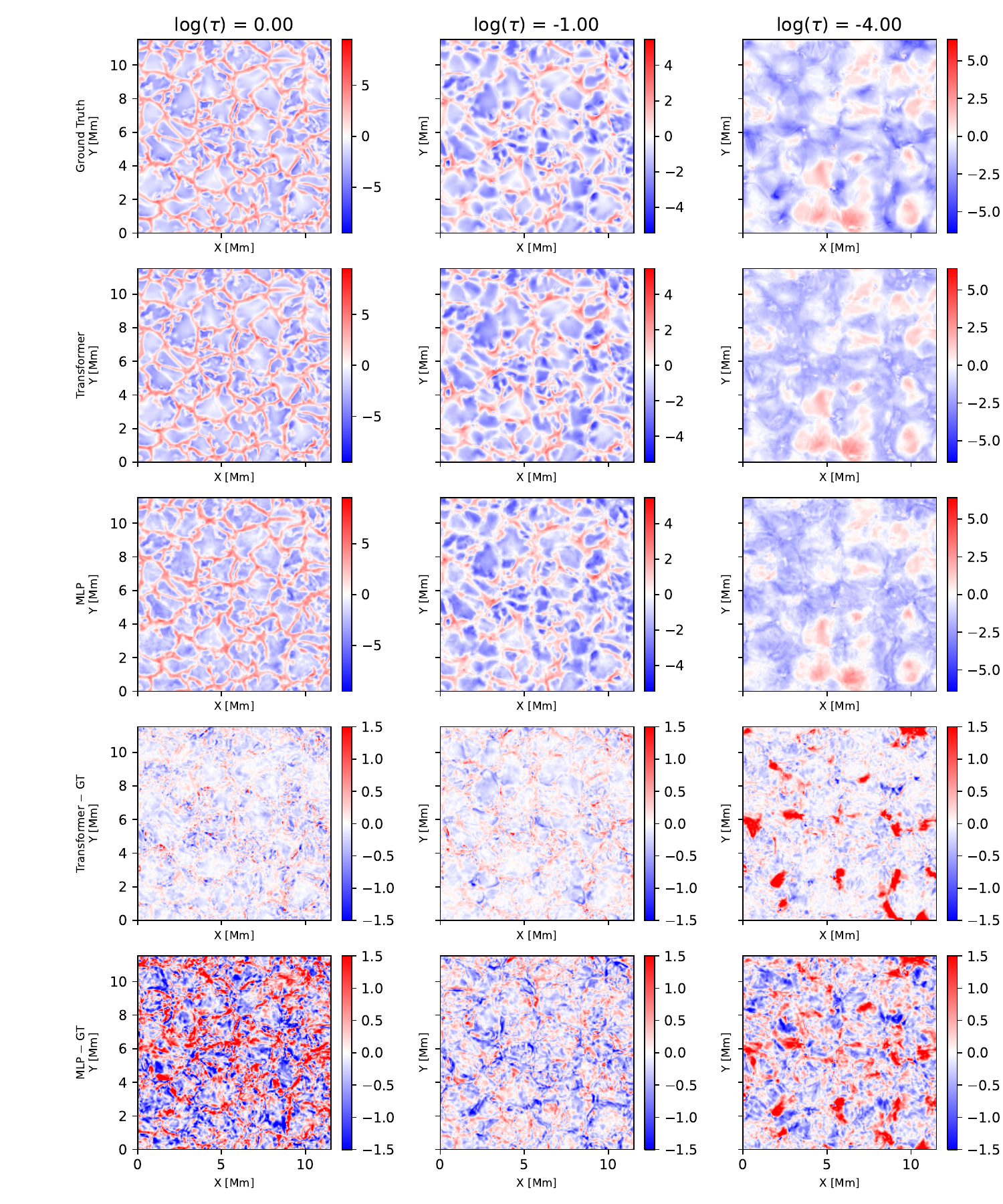}
    \caption{As in Fig.~\ref{fig:maps_T}, but for line-of-sight velocity. All units are km/s.}
    \label{fig:maps_vlos}
\end{figure*}

For both magnetic field strength (Fig.~\ref{fig:maps_B}) and line-of-sight velocity (Fig.~\ref{fig:maps_vlos}), the qualitative trends mirror those seen in the temperature maps. At lower optical depths ($\log(\tau) = 0$ and $-1$), the transformer accurately reconstructs the spatial morphology and small-scale features across the field of view, with residuals typically lower than those of the MLP. The transformer especially predicts a solution closer to ground truth for the weakest magnetic fields. The residual maps highlight the superior stability and consistency of the transformer. At $\log(\tau) = -4$, both models show larger and more systematic deviations, but the transformer continues to better reproduce the overall morphology. For magnetic field strengths at higher layers both models noticeably struggle more than in the photosphere. The transformer more successfully predicts the strongest magnetic field strengths at the higher layer, but also fails to predict the largest up-flows.

\begin{figure*}
    \centering
    \includegraphics[width=\textwidth]{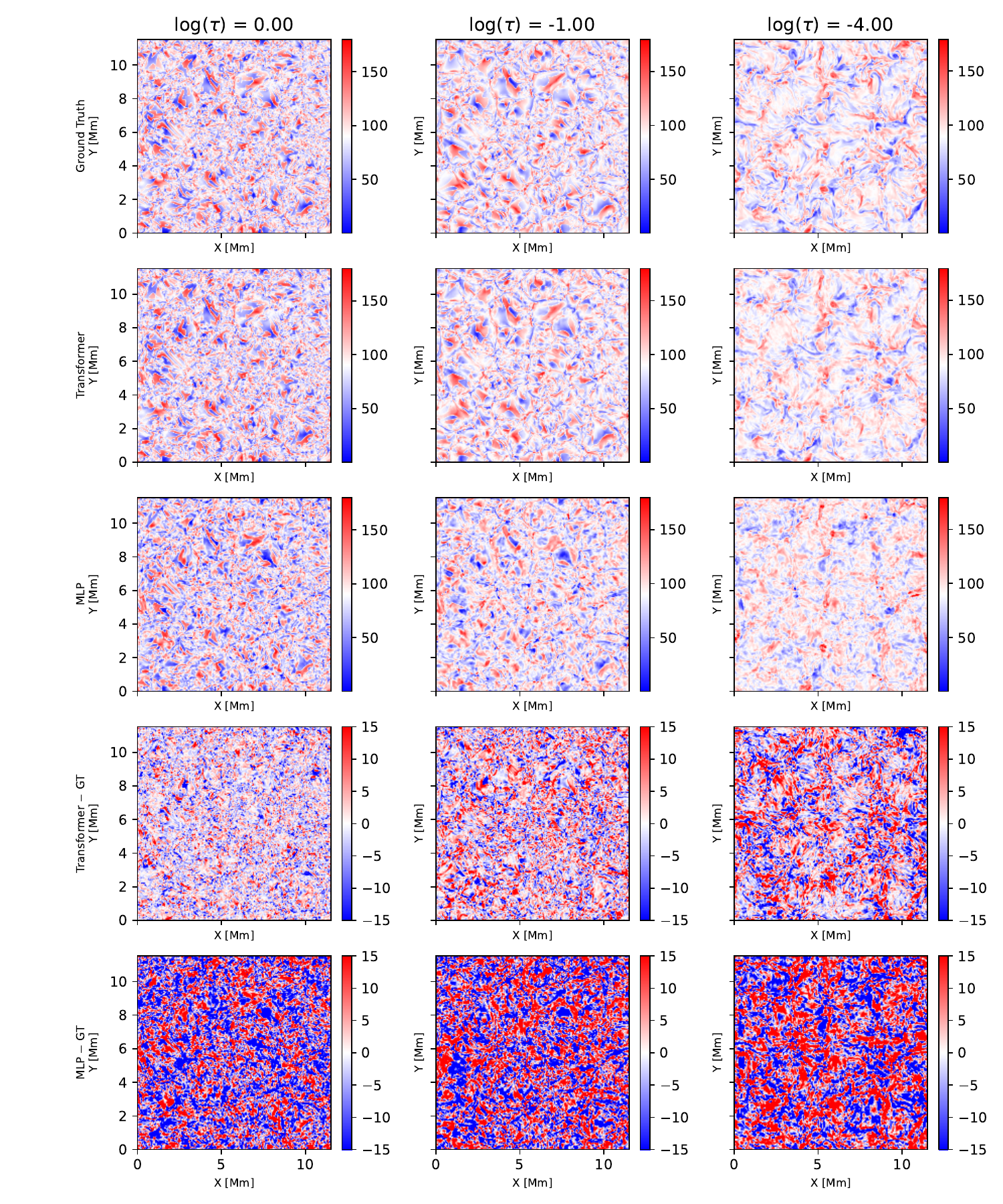}
    \caption{As in Fig.~\ref{fig:maps_T}, but for inclination. All units are degrees.}
    \label{fig:maps_incl}
\end{figure*}

For inclination (Fig.~\ref{fig:maps_incl}), the transformer is a clear leader in terms of reproducing the highly complex small-scale morphological features at all three optical depths. 

\begin{figure*}
\includegraphics[width=\textwidth]{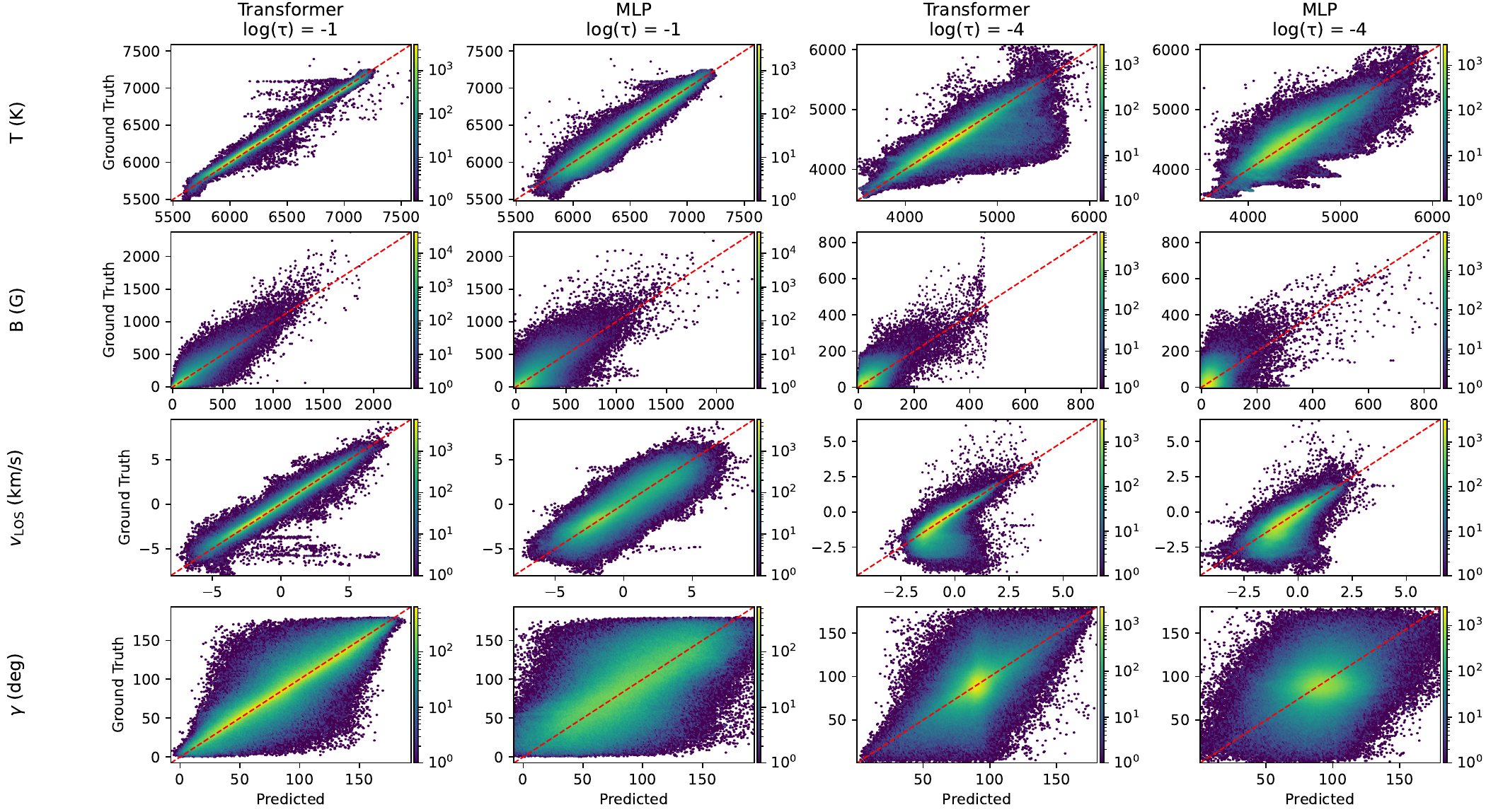}
\caption{
    2D histograms comparing predicted and ground truth atmospheric parameters at two representative optical depths (\(\log(\tau) = -1\) and \(\log(\tau) = -4\)) for both the transformer and MLP models. Each row corresponds to a different atmospheric parameter (from top to bottom: temperature, magnetic field strength, line-of-sight velocity, and magnetic inclination). The colour scale (i.e. pixel density) is on a logarithmic scale. The dashed red line indicates the one-to-one correspondence.}
\label{fig:histograms}
\end{figure*}

The comparative performance of the two models is further illustrated in Figure~\ref{fig:histograms}, which presents 2D histograms of the predicted versus ground truth values at two representative optical depths for all pixels in the test snapshot, with the density of points indicated by a logarithmic colour scale. This means there are order(s) of magnitude more pixels along the brightest regions of the histograms. The dashed red line corresponds to perfect one-to-one agreement. For all parameters at both depths, the transformer predictions cluster more tightly along the one-to-one line, indicating not only high accuracy but also a strong absence of systematic bias. In contrast, the MLP results show greater scatter in all parameters and both depths. For inclination in the photosphere, the transformer has a much higher density of correct predictions along the one-to-one correspondence line, while the MLP is more broadly dispersed. This pattern is also evident for temperature and line-of-sight velocity. This confirms that the transformer model delivers more reliable and stable inversions across atmospheric heights, however both models show the greatest differences emerging towards the chromospheric regime.

\begin{figure}
    \centering
    \includegraphics[width=1.0\linewidth]{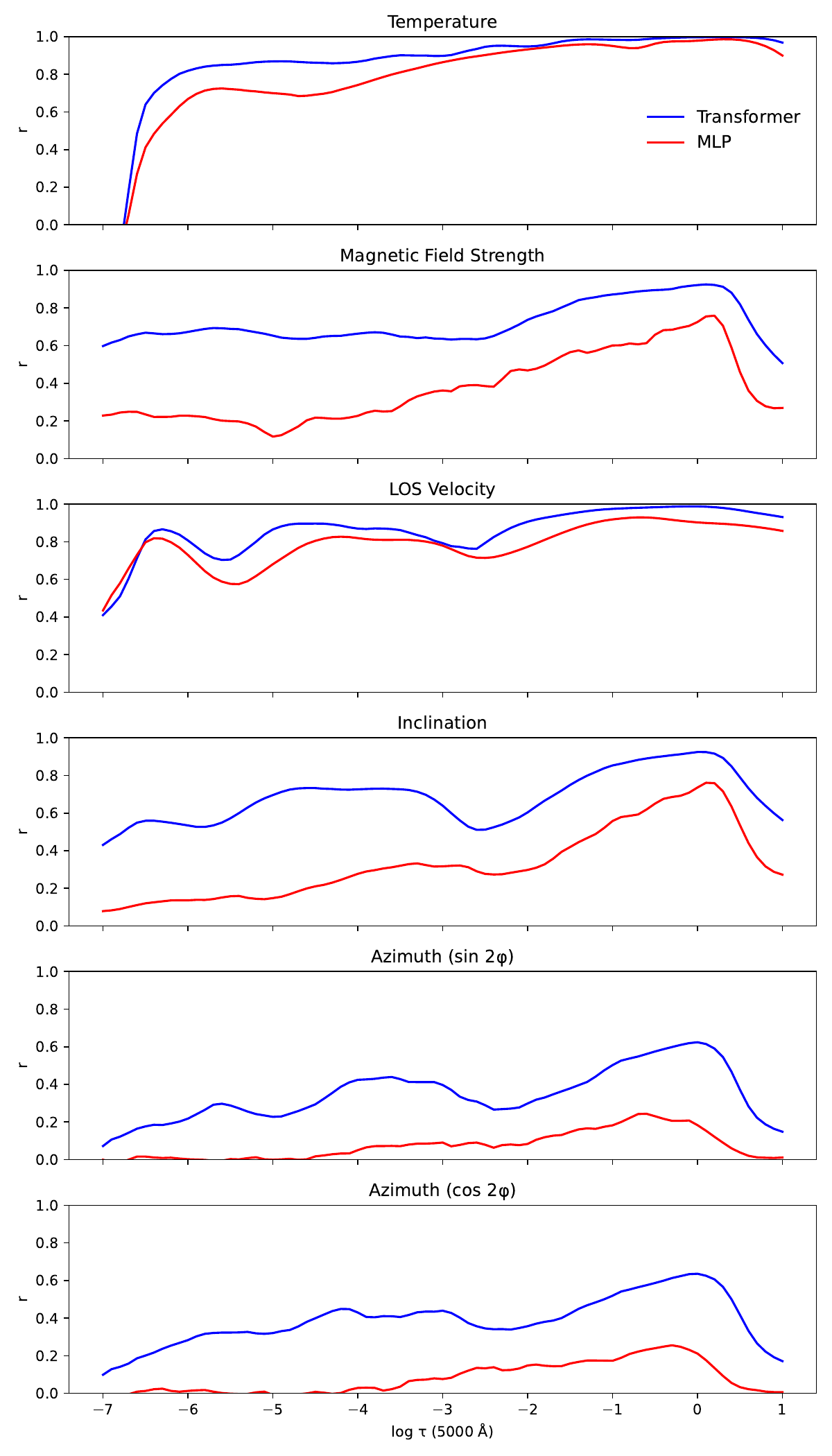}
    \caption{Correlations between predicted and ground truth values as a function of optical depth, for all inverted atmospheric parameters. Each panel shows the Pearson correlation coefficient ($r$, solid lines) and the Spearman rank correlation coefficient ($\rho$, dashed lines) evaluated at each depth, for both the transformer (blue) and MLP (red) models. The correlations are computed over all pixels in the test snapshot for each optical depth point.}
    \label{fig:correlations_TvsMLP}
\end{figure}

Correlations are not useful for examining the true error between the predictions of the transformer and MLP and the ground truth, but they are useful as a measure of performance because if one model is outperforming the other its correlations should be significantly higher. Figure \ref{fig:correlations_TvsMLP} shows the Pearson correlation coefficients, $r$, as a function of optical depth for the transformer and the MLP for each parameter. One can observe that both the transformer and MLP achieve similar highly positively correlated values ($r$ close to $1$) in the photosphere for temperature and line-of-sight velocity. For temperature, the $r$ drops for the MLP above $\log(\tau) = -3$, while the transformer remains stable and highly correlated. At even higher layers, the $r$ remains above 0.8 for the transformer in temperature between $\log(\tau) = -3$ and $\log(\tau) = -6$.

For the magnetic parameters, including $B$, $\gamma$, and $\phi$, the transformer achieves significantly higher correlations throughout the entire atmosphere. The significant improvement of the transformer over the MLP at predicting parameters describing the strength and orientation of the magnetic vector is clearly demonstrated.

The $r$ for inclination, azimuthal parameters, and to a lesser extent the line-of-sight velocity, shows a dip between $\log(\tau) = -2$ and $\log(\tau) = -3$. This may represent a gap in sensitivity between the chromospheric Ca II lines and the Fe I photospheric lines. One may have expected the upper photospheric lines to fill this gap, but the very weak polarisation signals in these lines is a limiting factor.

\subsection{Impact of Gaussian noise on inversion performance}
In any practical deployment of machine learning for spectropolarimetric inversion, the model must be trained under conditions that realistically reflect the characteristics of the observational data it will encounter. Real spectropolarimetric observations of the Sun are always affected by photon noise, sometimes approximated as uncorrelated, zero-mean Gaussian noise with a standard deviation estimated from the continuum of the Stokes profiles. Therefore, to ensure the model’s performance is representative of real-world usage, we add Gaussian noise to the synthetic training profiles at the same amplitude as would be measured in the continuum of actual observations. At inference, we apply the same noise level to unseen profiles, mimicking the noise conditions of real data. This approach ensures that the model both learns to handle the target SNR regime and is evaluated under precisely the same constraints it will face in application.

\begin{figure}
    \centering
    \includegraphics[width=1.0\linewidth]{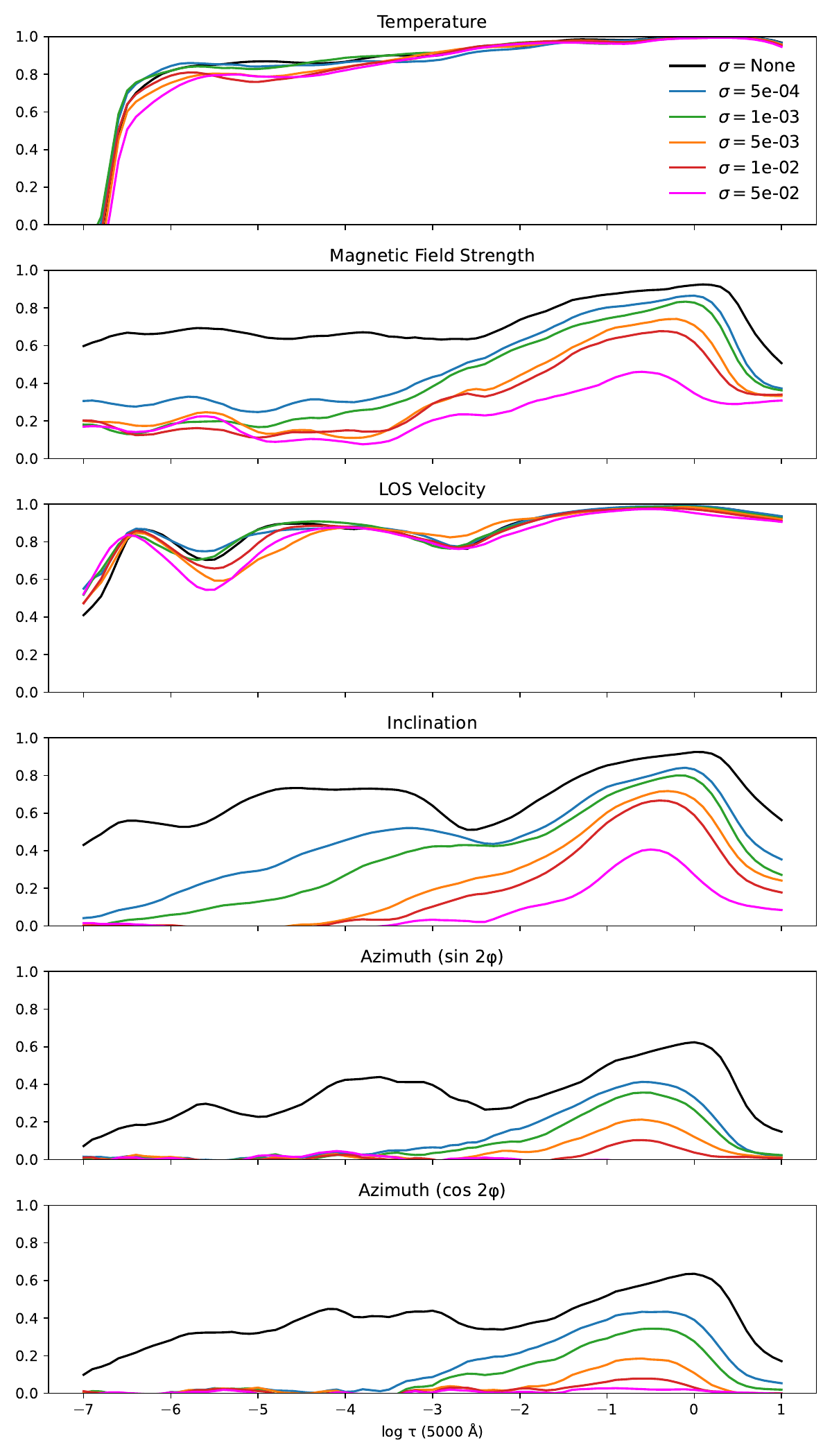}
    \caption{As in Fig.~\ref{fig:correlations_TvsMLP}, but for the Pearson correlation coefficient for transformer models trained and evaluated at different Gaussian noise levels. The black lines correspond to a model trained and evaluated without added noise. Coloured curves show results for models trained and evaluated with matched Gaussian noise amplitudes added to the Stokes profiles, at standard deviations ($\sigma$) indicated in the legend.}
    \label{fig:correlations_noise}
\end{figure}

Figure \ref{fig:correlations_noise} shows the $r$ as a function of optical depth for the transformer when trained at different noise levels, ranging from $5\times10^{-4}$~I$_c$ to $5\times10^{-2}$~I$_c$. The noise degradation has no significant impact on the inference of temperature between $\log(\tau) = 1$ and $\log(\tau) = -4$. In the most extreme noise cases, some minor degradation in the $r$ value begins at around $\log(\tau) = -4$, and all noise levels rapidly deteriorate at $\log(\tau) = -6$. The inference of line-of-sight velocities only begins to degrade above $\log(\tau) = -4$ for the higher noise values, but the two lowest noise levels never significantly fall below the noiseless benchmark.

The impact on the retrieval of the information describing the magnetic vector is much clearer and is consistent at all optical depths. There is a very clear trend in $B$, $\gamma$, and $\phi$ showing that as the standard deviation of the Gaussian noise increases, the correlation between the predicted parameters and the ground truth degrades throughout the atmosphere.

\subsection{Transformers as a regulariser}
\begin{figure*}
    \includegraphics[width=\textwidth]{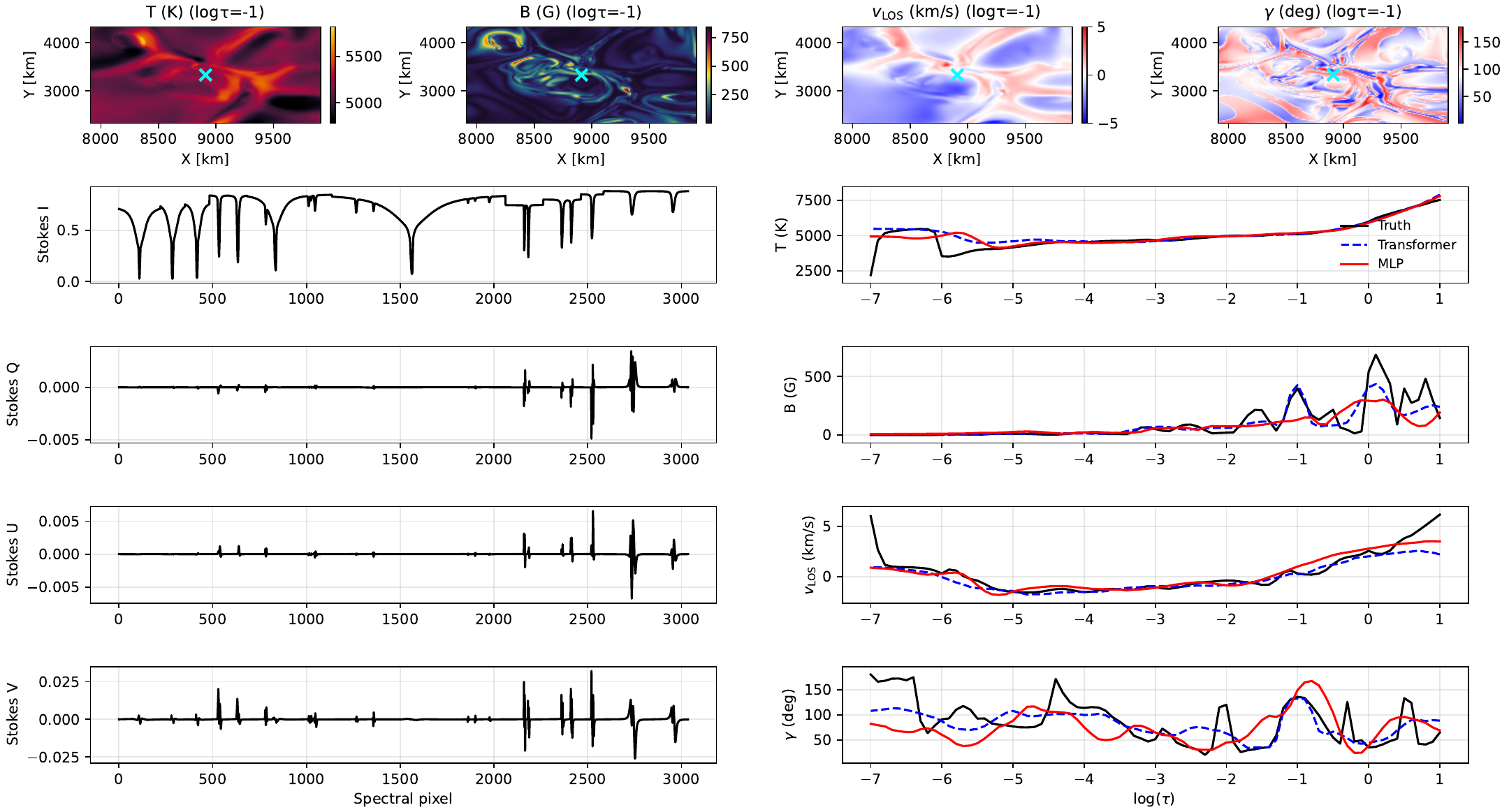}
    \caption{Comparison of ground truth atmospheric parameters transformer and MLP predictions for a pixel near an intergranular-granule boundary. The upper row shows a close-up view of $T$, $B$, $v_\mathrm{LOS}$, and $\gamma$ ground truth values at $\log(\tau) = -1$. Shown in the left column is the Stokes vector produced by DeSIRe from the ground truth atmospheric model, for the selected pixel, as a function of spectral pixel index. Shown in the right column is the atmospheric parameters as a function of optical depth. The black curve shows the ground truth, the dashed blue curve shows the transformer prediction, and the red curve shows the MLP prediction. }
    \label{fig:example_pixel}
\end{figure*}

To qualitatively demonstrate the performance of both ML models, we selected a pixel near the intergranular-granule boundary and compared the predicted atmospheric parameters and the corresponding synthetic Stokes profiles. Figure~\ref{fig:example_pixel} presents the ground truth atmospheric model parameters alongside the MLP and transformer predictions as a function of optical depth, as well as the  Stokes $I$, $Q$, $U$, and $V$ profiles produced from the ground truth model - the MLP and transformer have predicted the model parameters from this Stokes vector. The Stokes vector contains very weak chromospheric polarisation signals, and the MLP and transformer both correctly predict very low $B$ values in the chromosphere. For temperature, both models make excellent predictions. For line-of-sight velocity, the ground truth has a clear gradient in optical depth which both models correctly predict. For the two magnetic parameters shown, inclination and magnetic field strength, the transformer prediction matches the stratification of the ground truth in the photosphere much more closely.

\begin{figure*}
    \includegraphics[width=\textwidth]{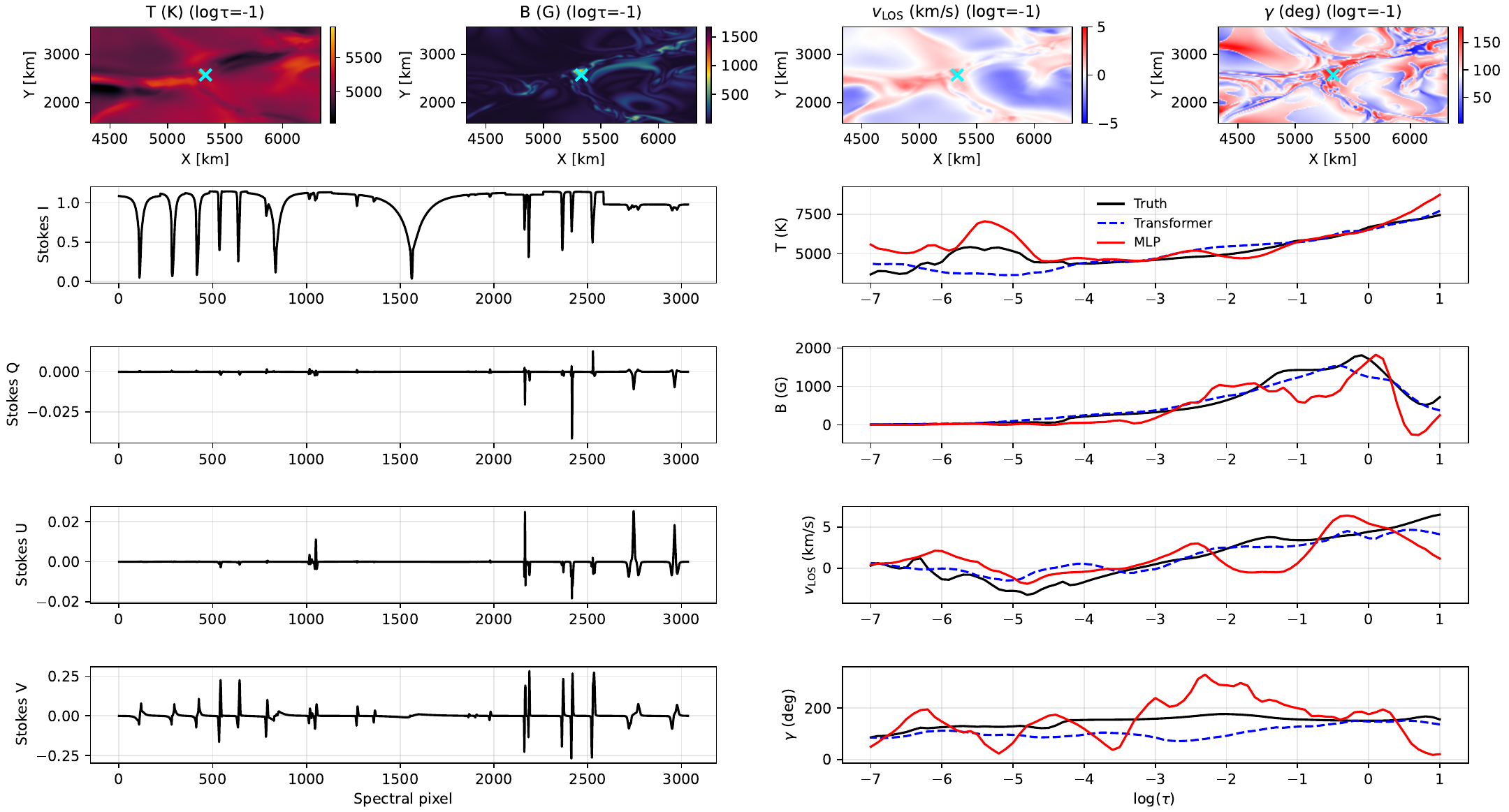}
    \caption{As in Fig.~\ref{fig:example_pixel}, but for a pixel in a relatively strong magnetic concentration located in an intergranular lane.}
    \label{fig:example_pixel_MBP}
\end{figure*}

Figure~\ref{fig:example_pixel_MBP} shows inference for a pixel in a relatively strong magnetic concentration located in an intergranular lane. The Stokes vector for this model has larger circular polarisation amplitudes even in the upper photospheric and lower chromospheric lines. As in the previous case, the transformer and MLP both track the ground truth temperature closely in the photosphere, however above this both ML predictions deviate significantly from the true values. The same is true for line-of-sight velocity, although the transformer has a lower residual error even in the photosphere. For magnetic field strength the transformer clearly predicts more accurately than the MLP. Consistently with the previous example, the MLP struggles significantly when predicting inclination, even predicting unphysical values (both above $180^\circ$ in this case) while the transformer prediction stays physically bounded, even though it has a significant error above the photosphere.

\subsection{Interpretability of cross-attention}
To explore how the transformer achieves its performance, we inspected the cross-attention between optical depth queries and the common encoder output (spectral tokens). In the decoder, each depth query $q_i$ is compared with all wavelength keys $k_j$ through the scaled dot-product attention mechanism (Eq.~\ref{eqn:attn}), producing a weight $a_{ij}$ that expresses how strongly optical depth $i$ attends to wavelength $j$. The resulting attention matrix $A \in \mathbb{R}^{n_\tau \times n_\lambda}$ can be visualised as a heatmap across wavelength and depth, showing, for each decoder layer and head, how strongly each depth query attends to each wavelength. Here we retrieve the cross-attention from the last decoder layer because it provides the final cross-attention signal; after the layer’s feed-forward sublayer, the resulting decoder states are passed to the output head and thus most directly influence the predictions. Maps can be produced for each head.

\begin{figure*}
    \includegraphics[width=\linewidth]{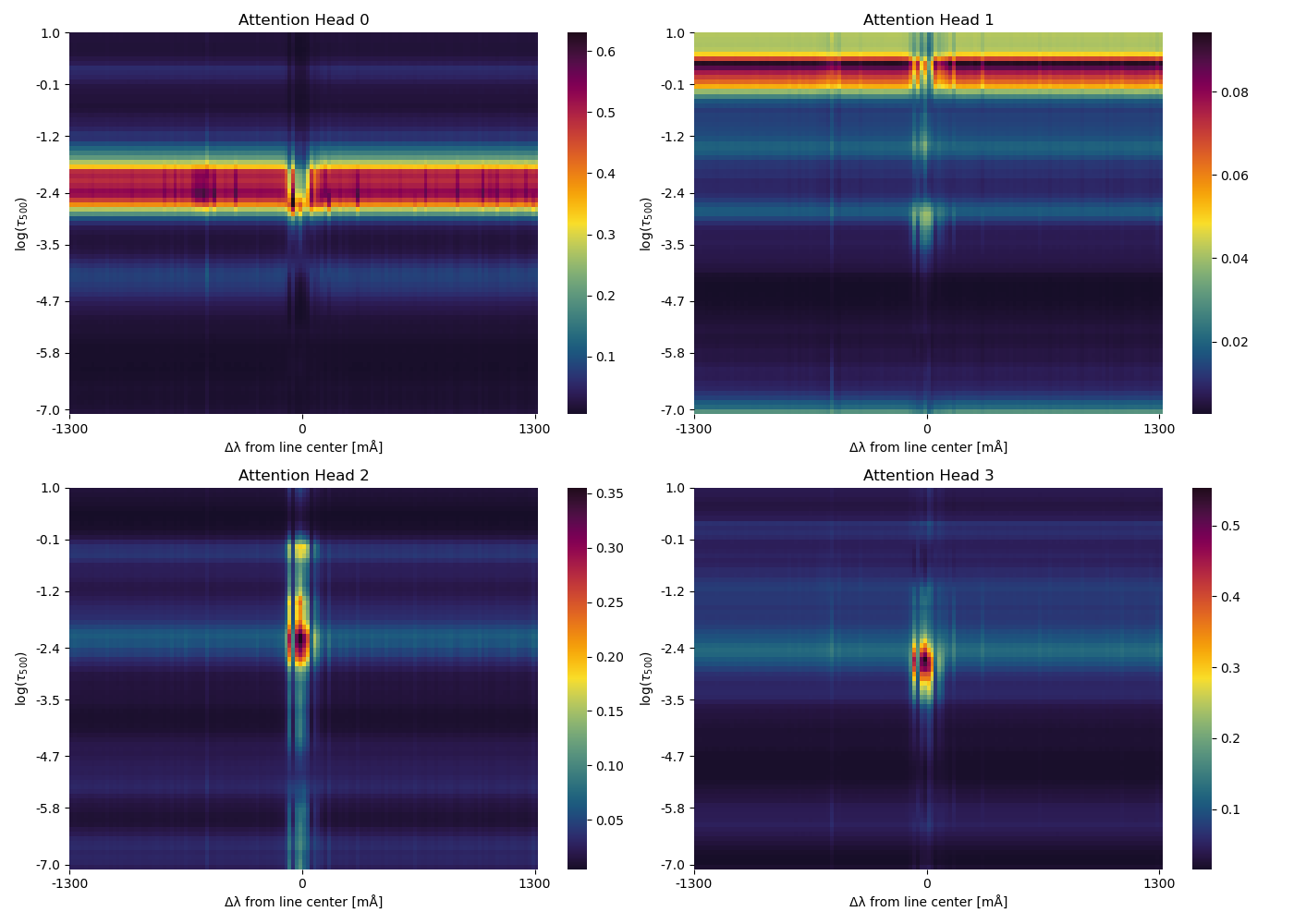}
    \caption{Cross-attention maps for the Na~I~D$_1$ line retrieved from the final decoder layer, shown separately for each of the four attention heads. Each panel shows the row-normalised attention weight assigned between spectral tokens and depth queries. The maps are computed by averaging over $10000$ Stokes vectors. Each row is scaled by its global maximum; values indicate which wavelengths are preferred at that depth, not how much absolute attention is assigned.}\label{fig:attn}
\end{figure*}

We find that each of the four attention heads specialises on different spectral regions and depth ranges. This is illustrated in the sample attention maps shown in Fig.~\ref{fig:attn} for the Na~I~D$_1$ line. For visual comparison with response functions, we row-normalise the attention maps across all wavelengths (i.e. $3036$~wavelengths, including those outside the Na~I~D$_1$ line) per head so that each depth's most attended wavelength is scaled to unity; this way we can clearly see which wavelength stands out most and how all the others compare to it, without the plot being skewed by differences between depths or heads. For this line, one head concentrates on the wings at photospheric depths (log~$\tau \approx -1$ to $-2.5$), while another attends primarily to the continuum-forming layers (log~$\tau \approx 0$), and while others focus on the line cores at higher layers (up to log~$\tau \approx -3.5$). Note that because the maps are row-normalised, they highlight qualitative patterns of focus within each depth and head, rather than giving absolute attention strengths that can be compared across depths or heads. Similar qualitative behaviour is observed for other lines, especially all the upper photospheric lines.

\begin{figure*}
    \includegraphics[width=\linewidth]{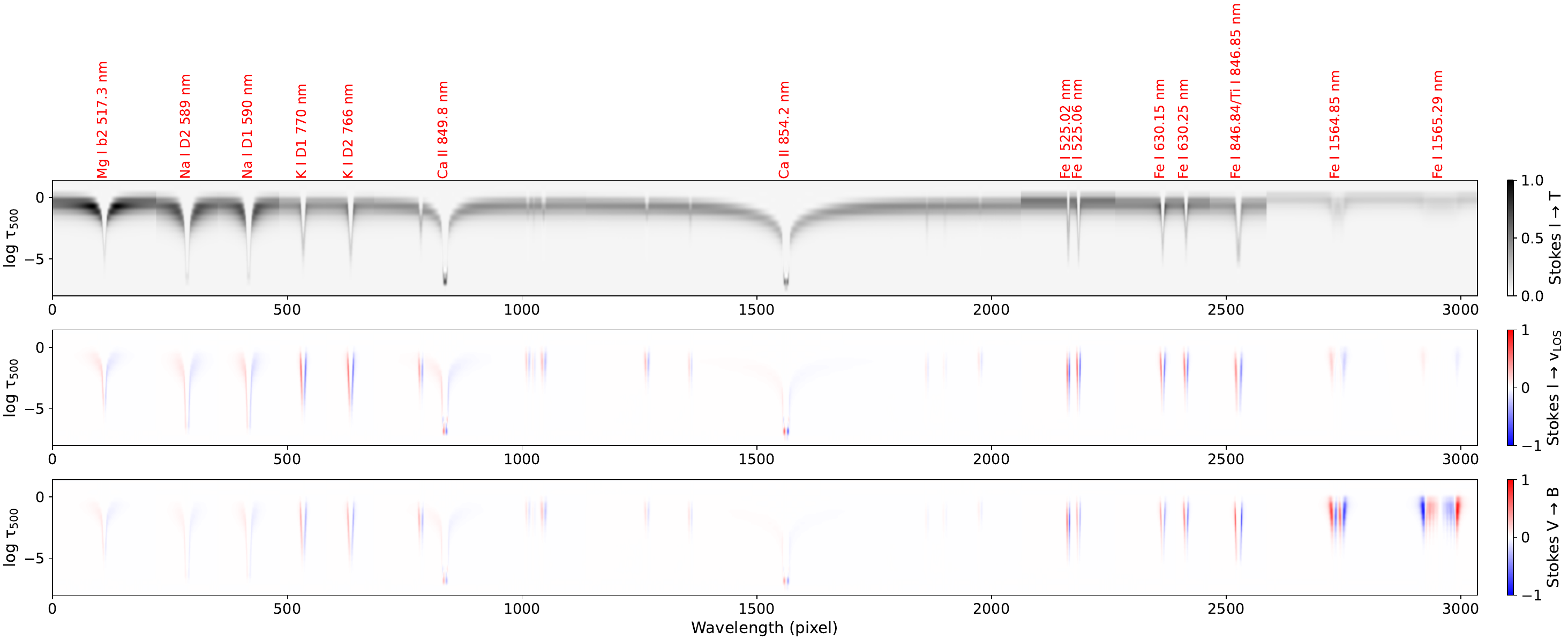}
    \caption{Response functions of Stokes $I$ to $T$ (upper panel), Stokes $I$ to $v_\mathrm{LOS}$ (middle panel), and Stokes $V$ to $B$ (lower panel), for all spectral lines included in the input sequence, computed from a semi-empirical quiet-Sun reference model with $B = 500$~G, $v_\mathrm{LOS} = 0.5$ km/s and $\gamma = 45^\circ$. The response functions have been normalised by the maximum value for visualisation.}\label{fig:RFs}
\end{figure*}

This behaviour is qualitatively consistent with the response functions of these lines; wings are most sensitive to photospheric conditions, while cores are sensitive to higher layers, as demonstrated in Figure~\ref{fig:RFs} for the response function of Stokes $I$ to $T$. This is interpreted as evidence that the transformer assigns weight to spectral regions where the line is known to carry diagnostic information.

Because we jointly predict six atmospheric parameters and attend over per-wavelength tokens that embed all four Stokes components together, this comparison is not one-to-one with parameter-specific response functions. Nevertheless, the alignment between attention maps and known line-formation properties suggests that the transformer has learned to extract information from physically diagnostic parts of the spectrum. 

It is important to note that the transformer also contains feed-forward networks in each encoder and decoder layer, so these cross-attention maps should not be over-interpreted as a complete explanation of the model's learned behaviour. Rather, they are qualitative indicators of which spectral regions and depths are emphasized during the mapping from Stokes profiles to atmospheric parameters.

\section{Discussion}
In this proof of concept study, we have demonstrated the application of a transformer-based neural network, originally developed for natural language processing, to the inversion problem of inferring stratified atmospheric parameters from full-Stokes, multi-line spectropolarimetric data generated from realistic 3D MHD simulations. Our results show that the transformer model outperforms an MLP baseline in nearly all aspects: reproducing the spatial and vertical morphology of temperature, magnetic field strength, and velocity; achieving higher correlations with the ground truth at all optical depths; and, crucially, delivering physically plausible solutions even in challenging chromospheric layers. The best-performing MLP has approximately $55.2\,\mathrm{M}$ trainable parameters, whereas the winning Transformer has about $5.42\,\mathrm{M}$. Despite being roughly ten times smaller, the Transformer outperforms the MLP. The MLP is parameter-heavy but relatively compute-light, while the Transformer is parameter-light but compute-heavy.

One of the most significant findings is the improvement of the transformer’s predictions in the magnetic parameters. While the MLP baseline performs at least comparably to the transformer in the deeper atmospheric layers in temperature and velocity, the transformer is markedly more robust at accurately capturing the fine structure of both granules and intergranular lanes, clearly outperforming the MLP in the inference of magnetic parameters, even where the polarisation signal is weak. Positional embedding allows the transformer to remain regularised in its predictions even in these challenging regimes, while the MLP violates physically plausible bounds in its predictions in inclination, and we have verified this is not an isolated incident for the MLP.

The attention mechanism at the core of the transformer architecture enables the model to leverage non-local spectral features, potentially corresponding to Zeeman splitting or subtle asymmetries, which may be overlooked by models that treat each wavelength as just another input feature, with no explicit bias toward local or long-range spectral relationships. Although a detailed interpretability study is beyond the current scope, initial inspection of the attention maps suggests that the model prioritises physically meaningful regions of the spectrum for inference. This could open new avenues for understanding how neural networks extract physical parameters from complex solar spectra.

Despite these promising results, several limitations and caveats should be acknowledged explicitly. First, the models are trained and evaluated exclusively on synthetic data produced by the MANCHA simulation code. While this simulation is one of the most realistic options for quiet Sun conditions due to its inclusion of the Biermann battery effect, its chromosphere is not expected to be realistic. The “ground truth” in this context is therefore only as realistic as the underlying simulation. Extension to simulations which might provide a more realistic chromosphere, or to training sets that combine multiple simulation codes, will be necessary to ensure that the model can generalise to a broader set of solar conditions. However, expanding the training set to include models and Stokes vectors from multiple simulation codes with differing physical assumptions risks introducing degenerate mappings between Stokes profiles and atmospheric parameters. This can hinder model convergence and reduce generalisation, unless carefully controlled. In particular, differences in chromospheric realism and magnetisation between simulations may lead to conflicting training signals that cannot be reconciled by the model.

Furthermore, we do not include instrumental effects, such as spatial or spectral point spread functions (PSF) or stray light, which are present in real observations. Noise was added to the synthetic profiles to better mimic realistic conditions, and our analysis shows that moderate noise levels have little effect on temperature and velocity inference in the photosphere and lower chromosphere, but begin to degrade the inference of magnetic parameters at all optical depths as expected. However, the noise applied here is idealised (Gaussian and uncorrelated), whereas real observations often feature more complex and structured noise properties.

In terms of physical representation, the sin(2$\phi$) and cos(2$\phi$) parameterisation alleviates the discontinuity problem but does not resolve the ambiguity. While this representation improves numerical stability, future work could explore alternative approaches to capturing directional ambiguities, especially when working with real observations of the Sun. Of course, one could choose not to infer the azimuth if it is not a priority. Azimuthal predictions are discussed in Section~\ref{section:appendix}.

The current approach inverts each pixel independently, without explicit consideration of spatial context, just like traditional one-dimensional inversion codes. While this allows for highly parallelisable processing and serves as a proof of concept, incorporating spatial relationships, whether through convolutional layers or regularisation techniques, could further improve performance.

\section{Outlook}

This study provides the first transformer-based model for full-vector Stokes inversions in solar physics, enabling accurate and efficient inference of atmospheric parameters over many spectral lines and at many optical depths. We anticipate a follow-up study to apply the transformer to real spectropolarimetric observations, such as those from Sunrise III, DKIST, and other current or next-generation facilities. Training on simulations with advanced chromospheric physics will be critical to ensure robust generalisation. An open question is whether transformers might be useful in providing initial guesses for subsequent inversions, though this requires careful testing in future work.

\begin{acknowledgments}
We thank the anonymous referee who helped us significantly improve the manuscript. R.J.C. would like to thank Robert Ryans and Ernst de Mooij for approving use of an NVIDIA Quadro RTX 8000 GPU for model development, as well as technical staff from the Kelvin2 HPC support teams in using their NVIDIA H100 GPUs. R.J.C. thanks Elena Khomenko for permission to use the MANCHA models. This is UKRI-supported research under the Science and Technology Facilities Council (STFC) grants ST/P000304/1 and ST/X000923/1. We acknowledge that the MANCHA simulations used in this paper are produced using the technical expertise and assistance provided by the Spanish Supercomputing Network (Red Española de Supercomputación), as well as the computer resources used: LaPalma Supercomputer, located at the Instituto de Astrofísica de Canarias, and MareNostrum based in Barcelona/Spain.

Data availability statement: The code that defined the models generated in this study is \hyperlink{www.github.com/r-j-campbell/SINN-inversions}{available on GitHub}. The MANCHA simulation code is \hyperlink{https://gitlab.com/Mancha3D/mancha}{available on their public repository}. The simulation data used in this study were generated by a third party and are not publicly archived. Access to these data may be available from the original authors on request.

\end{acknowledgments}

\begin{contribution}
R.J.C. conceived the project, designed and implemented the machine learning models, conducted the analysis and authored the paper. M.M. provided critical feedback on the framing of the study and the manuscript. C.Q.N. provided simulation data converted to an optical depth scale that formed the training and test set. All authors approved the manuscript before submission.

\end{contribution}

\software{PyTorch \citep{pytorch2019}, Matplotlib \citep{mpl}, Numpy \citep{numpy}, Astropy \citep{astropy3}.}

\appendix

\section{Azimuth predictions}\label{section:appendix}
\begin{figure*}
\includegraphics[width=0.95\textwidth]{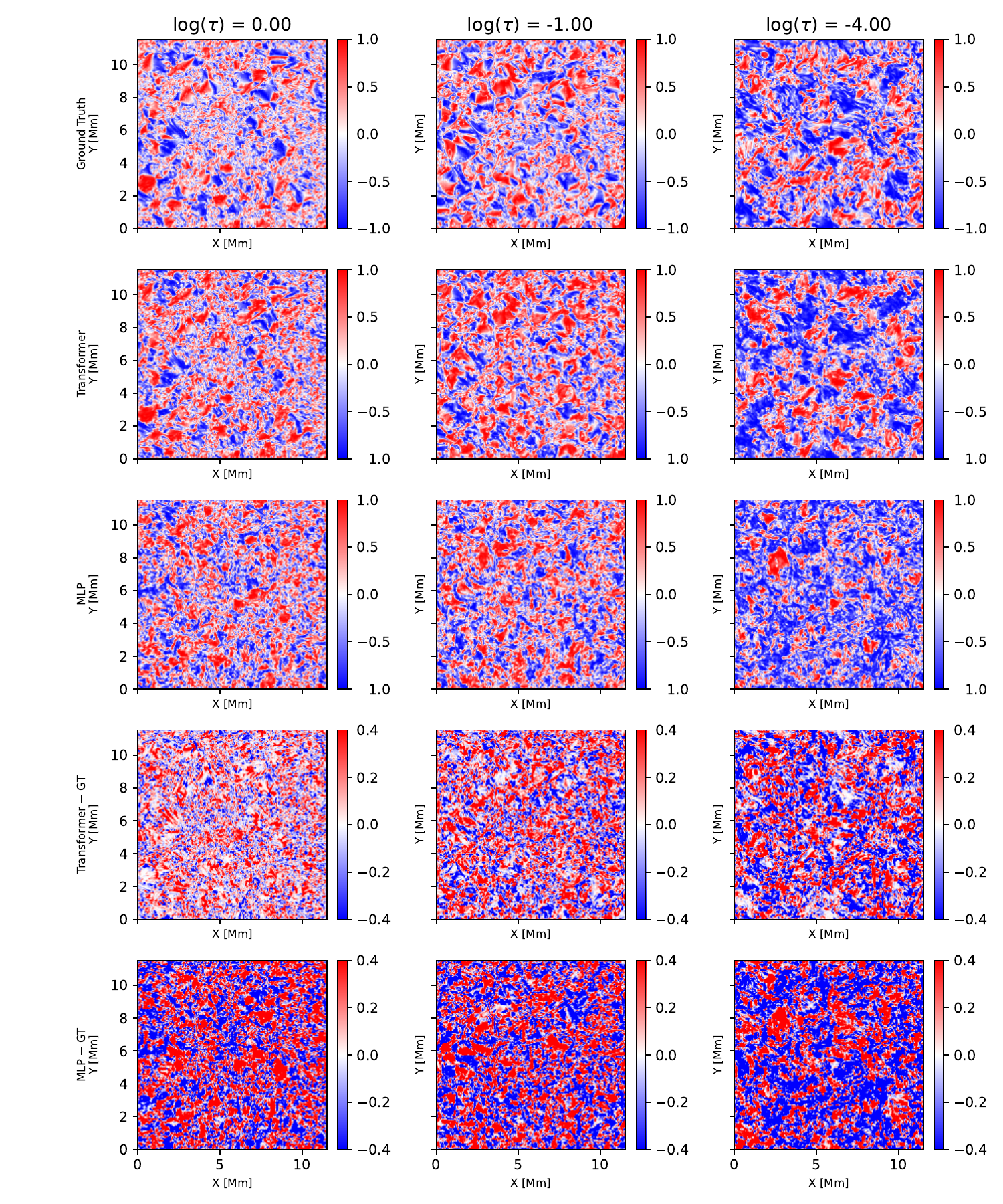}
\caption{As in Fig.~\ref{fig:maps_T}, but for $\sin(2\phi)$.}
\label{fig:sin_azi}
\end{figure*}

\begin{figure*}
\includegraphics[width=0.95\textwidth]{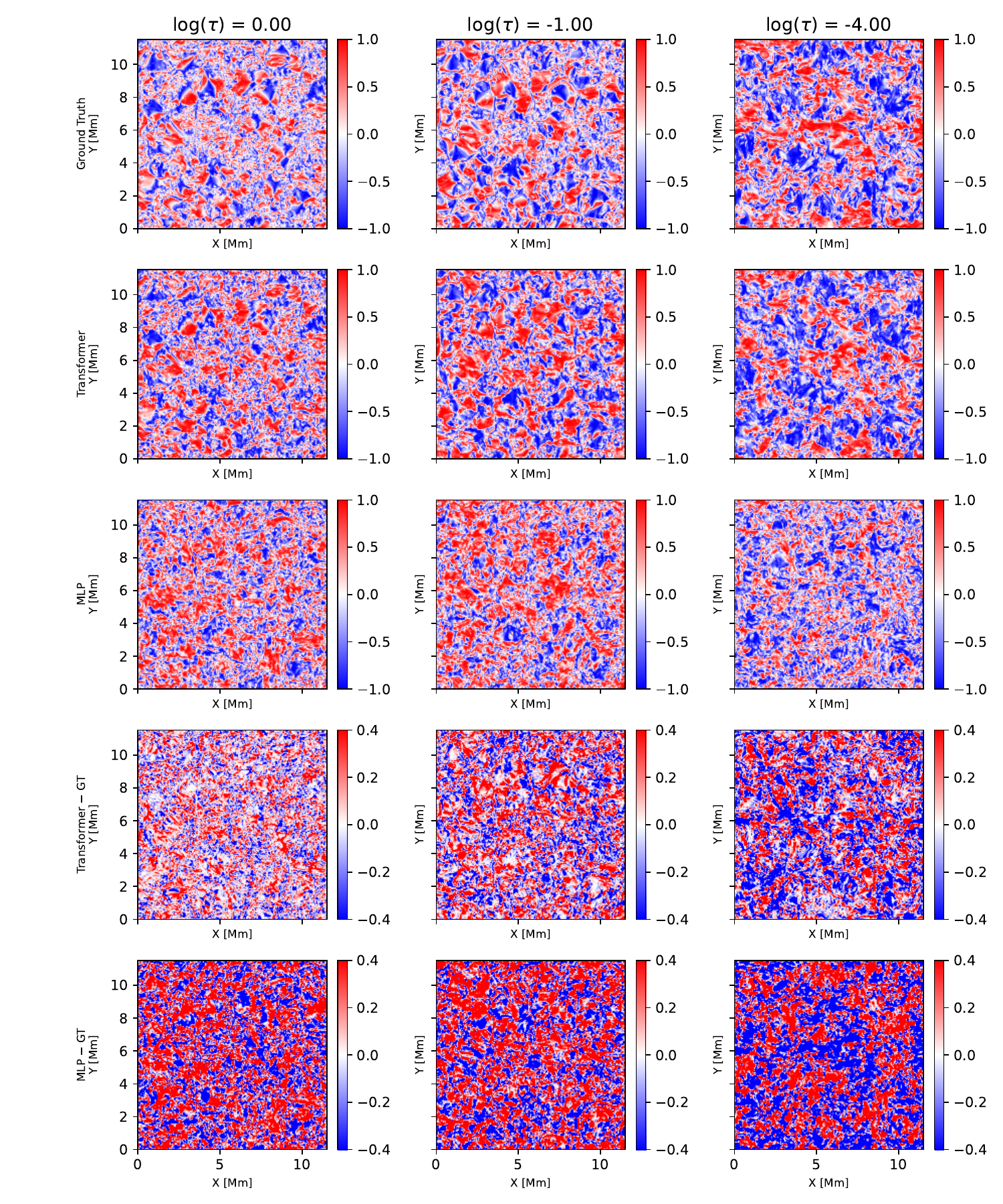}
\caption{As in Fig.~\ref{fig:maps_T}, but for $\cos(2\phi)$.}
\label{fig:cos_azi}
\end{figure*}

Figures~\ref{fig:sin_azi} and ~\ref{fig:cos_azi} show the azimuthal predictions for the MLP and transformer. In most of the snapshot, for both ML models, the azimuth has relatively large errors. The darkest regions in the residual maps do indicate however that there are distinct regions where the azimuth has been successfully predicted, and consistent with the other atmospheric parameters, the transformer clearly performs more effectively at this task than the MLP. In order for the azimuth to be constrained (without disambiguation), there must be sufficiently large polarisation amplitudes in Stokes $Q$ and $U$, which occurs in only a small minority of pixels due to the quiet Sun nature of the MANCHA snapshots.

\bibliography{sample701}{}
\bibliographystyle{aasjournalv7}

\end{document}